\documentclass[conference]{IEEEtran}
\IEEEoverridecommandlockouts
\usepackage{cite}
\usepackage{amsmath,amssymb,amsfonts}
\usepackage{algorithmic}
\usepackage{textcomp}
\usepackage{multirow}
\usepackage{threeparttable}
\usepackage{booktabs}
\usepackage{makecell}
\usepackage{graphicx}
\usepackage[hidelinks]{hyperref}
\usepackage{subcaption}
\usepackage[normalem]{ulem}
\usepackage{xcolor}
\usepackage{xspace}

\def\BibTeX{{\rm B\kern-.05em{\sc i\kern-.025em b}\kern-.08em
    T\kern-.1667em\lower.7ex\hbox{E}\kern-.125emX}}
\begin{document}

\newcommand{\codename}{\texttt{LOCAA}\xspace}

\newcommand{\wang}{\textcolor{purple}}
\newcommand{\MM}{\textcolor{blue}}
\title{\codename: An Agentic System for Automated Lossy Compressor Tuning
}

\author{
\IEEEauthorblockN{
Khondoker Mirazul Mumenin\IEEEauthorrefmark{1},
Dong Dai\IEEEauthorrefmark{2},
Sheng Di\IEEEauthorrefmark{3},
Franck Cappello\IEEEauthorrefmark{3},
Jinzhen Wang\IEEEauthorrefmark{1}
}
\IEEEauthorblockA{\IEEEauthorrefmark{1}University of North Carolina at Charlotte \emph{kmumenin@charlotte.edu, jwang96@charlotte.edu}}
\IEEEauthorblockA{\IEEEauthorrefmark{2}University of Delaware \emph{dai@udel.edu}}
\IEEEauthorblockA{\IEEEauthorrefmark{3}Argonne National Laboratory \emph{sdi1@anl.gov, cappello@mcs.anl.gov}}
}

\maketitle
\begin{abstract} Large-scale scientific simulations generate substantial data volumes, making lossy compression essential for reducing storage and data movement costs. However, users configure compressors through numerical error bounds (EBs) while often evaluating results using quality metrics and end-to-end performance. Because the relationship between an EB and these outcomes varies across datasets and compressors, identifying a suitable configuration typically requires exhaustive search, which can be time-consuming and computationally demanding. We present \codename, an Large language model-based scientific lossy compression auto-tuning agent that performs compression-in-the-loop search using tool-integrated execution, compressor-aware guidance, and persistent memory. {\codename supports user-defined objectives and constraints without requiring a specialized search strategy.} We evaluate \codename across three scientific applications, five compressors, and three use cases: {fixed-ratio compression}, {compression tuning under multiple quality constraints}, and {compression tuning under quality and time constraints}. For fixed-ratio search across twelve fields from two applications, \codename requires 1.98$\times$ fewer evaluation trials than binary search and 5.03$\times$ fewer than FRaZ on average. For compression ratio (CR) maximization under joint Peak Signal-to-Noise Ratio (PSNR) and Structural Similarity Index Measure (SSIM) constraints across six fields from the Community Earth System Model (CESM) application, \codename reduces the average evaluation trials from 61.5 using binary search to 17, corresponding to a 72.4\% reduction. Persistent memory further reduces the average number of trials by 27.9\% across multiple timesteps of the same field. These results demonstrate the potential of tool-augmented LLM agents to provide flexible and efficient compressor tuning across diverse scientific data and user-defined objectives. 
\end{abstract}

\section{Introduction}
Recent advances in computing have accelerated scientific discovery and enabled increasingly large-scale simulations. These simulations generate unprecedented volumes of multidimensional data. For example, large-scale NYX simulations can produce terabytes of data per snapshot~\cite{nyx2021}. Managing, storing, transferring, and analyzing data at this scale requires efficient data management and reduction techniques. Although larger simulations can improve spatial and temporal fidelity, storage and I/O limitations may force researchers to discard outputs or reduce their resolution, potentially removing critical physical features needed for subsequent analysis.

Scientific lossy compression techniques~\cite{di2016fast,SZ3,SZx,sperr} address this challenge by reducing data size while enforcing a user-defined EB. However, compression effectiveness depends strongly on the selected EB . Although the EB limits numerical reconstruction error, it does not directly translate to application-relevant quality measures such as  PSNR or  SSIM~\cite{wang2004ssim}, which may be more meaningful to domain scientists. Consequently, selecting an appropriate EB requires balancing data reduction against the quality preservation for the target analysis.

This disconnection creates a practical tuning problem. Because the mapping from an EB to the resulting quality is nonlinear and depends on both the dataset and the compressor~\cite{underwood2022optzconfig}, users typically resort to repeated compression, decompression, and metric evaluation to identify an acceptable configuration.

The problem is further compounded by the growing diversity of scientific lossy compressors and the increasing variety of scientific datasets. New compressors and compressor-family variants are often designed for specific data characteristics, accuracy requirements, or performance objectives. As a result, the most effective compression configuration can vary substantially across datasets and use cases, making manual tuning increasingly difficult.

Motivated by these challenges, we propose \codename, a Scientific \textbf{LO}ssy \textbf{C}ompression \textbf{A}uto-tuning \textbf{A}gent. \codename performs compression-in-the-loop search by iteratively selecting compressor configurations, evaluating their outcomes, and using the observed results to guide subsequent trials. The framework supports different user-defined objectives and quality constraints. \codename uses an large language model to reason over previous evaluations, adapt the search strategy, and identify effective compressor and EB configurations with fewer redundant trials, which is the key distinction compared with manual trial-and-error and existing search methods. The contributions of \codename are summarized as follows:

\begin{itemize}
\item We formulate compressor and error-bound selection as an agentic search problem and introduce \codename, which adapts its search strategy using measured compression feedback and user-defined constraints.

\item We develop a single-agent architecture that integrates compression-in-the-loop evaluation, task-specific search guidance, MCP-exposed tools, and lightweight persistent memory.
\item  We introduce compressor-aware searching strategies, including PSNR-driven initialization and staircase-aware navigation for compressors like ZFP, and quantify through ablation which components actually contribute to search efficiency.
\item We evaluate \codename across three applications, five compressors, and three use cases, showing it reaches feasible configurations using substantially fewer evaluation trials. 
\end{itemize}

The remainder of the paper is organized as follows. Section~\ref{sec:related} reviews related work. Section~\ref{sec:design} presents the design of \codename, followed by evaluation in Section~\ref{sec:eavl}. Section~\ref{sec:limitation} discusses limitations and future work, with conclusions in Section~\ref{sec:conclusion}.
\section{Related Work}
\label{sec:related}
\subsection{Large Language Models}
Large language models (LLMs) have progressed from conversational systems to the foundation of autonomous software agents~\cite{wang2024llmagents,yao2023react}. By combining language-model reasoning with external tools, memory, and iterative interaction, LLM agents can observe system state, reason about objectives, execute actions, and adapt to the resulting feedback. Unlike traditional chatbots that generate a response from a fixed conversational context, these agents can perform multi-step tasks requiring continuous interaction with external systems. This capability is now demonstrated by general-purpose commercial systems~\cite{openai2025deepresearch} and supported by frameworks for tool integration, workflow orchestration, memory, and multi-agent communication.

LLM agents have also been applied to scientific and systems tasks.  Curie~\cite{kon2025curie} automates scientific experimentation, ChemCrow~\cite{bran2024chemcrow} integrates language models with chemistry tools,  and STELLAR~\cite{Staller} uses LLM-driven autonomous reasoning to tune high-performance parallel file systems. These systems demonstrate that LLM agents can combine domain knowledge, computational tools, and iterative feedback to address complex scientific workflows.

Scientific lossy compression tuning has similar characteristics. The agent must interpret user-defined objectives and quality constraints, select a compressor and EB, invoke compression and evaluation tools, and revise subsequent decisions using observed CR, reconstruction quality, and runtime. These requirements motivate the use of an LLM agent as an adaptive search mechanism for identifying effective compressor configurations.

\subsection{Error-Bounded Lossy Compression}

Error-bounded lossy compression reduces the storage and data movement costs of scientific datasets while enforcing user-specified accuracy constraints. Based on how they de-correlate the input, existing compressors are broadly prediction-based or transform-based.

Prediction-based compressors exploit spatial or temporal redundancy by estimating each value from neighboring samples. SZ2~\cite{di2016fast} quantizes the prediction error and applies entropy and lossless coding to the resulting codes. SZ3~\cite{SZ3} extends this with a modular architecture whose prediction, quantization, encoding, and lossless-compression stages are selected independently, adapting the pipeline to a given dataset. SZx~\cite{SZx} instead emphasizes throughput via lightweight, CPU/GPU-specific operations while remaining competitive on CR.

Transform-based compressors map the data into a representation where redundancy is encoded more effectively, then compress the coefficients with specialized coding. ZFP~\cite{Lindstrom2014} applies a non-orthogonal block transform and encodes coefficients progressively using an embedded bit-plane representation, supporting scalable execution and random access to independently compressed regions. SPERR~\cite{sperr} applies a bi-orthogonal wavelet transform and a SPECK-inspired coding scheme with outlier handling for point-wise error constraints, giving scalable execution under strict tolerances.

\paragraph{Error-Bound Search}
Selecting an EB that meets user requirements is an iterative configuration problem: each iteration compresses and decompresses with a candidate EB, and the resulting CR and quality metrics guide the next candidate. {Prior work reduces direct evaluations by predicting these outcomes}, differing mainly in how the input and compression process are characterized. Sampling-based approaches estimate behavior from a data subset~\cite{8821342}, with accuracy tied to how well the sampling reflects the compressor's internal processing. Feature-based approaches build statistical or spatial descriptors~\cite{underwood2023black,krasowska2021exploring} and map them to outcomes via regression or neural networks. Most target CR; Jin et al.~\cite{jin2022improving} add PSNR and SSIM, and learning-based methods~\cite{mumenin2024qualitynet} predict PSNR, SSIM, and autocorrelation error directly from the input, but require separate training per dataset, compressor, and metric. All such methods yield estimates rather than measured outcomes, so accuracy degrades under distribution shifts or unseen compressor behavior, which is especially risky under strict fidelity constraints where a bad estimate can violate the requested quality.
Direct-search approaches instead evaluate candidates through actual executions. Binary search works when the EB-metric relationship is monotonic, but responses may be non-monotonic, discontinuous, or staircase-shaped~\cite{underwood2020fraz}. FRaZ uses derivative-free global optimization to hit a target compression ratio~\cite{underwood2020fraz}, and OptZConfig generalizes this to user-defined fidelity metrics and multiple parameters with early termination and parallel evaluation~\cite{underwood2022optzconfig}.

Our method also evaluates every candidate EB through actual compression and decompression, but replaces the fixed search algorithm with a tool-augmented LLM agent that interprets the measured metrics and chooses the next action, avoiding surrogate error while supporting diverse compressors, metrics, and objectives.

\subsection{Motivation}
Compression tuning requires repeatedly evaluating candidate EBs and measuring the resulting quality metrics. The effectiveness of a search strategy depends strongly on how these metrics respond to changes in the EB. Conventional strategies such as binary search are efficient when the response is monotonic but become unreliable when it is irregular or non-monotonic. Derivative-free methods such as FRaZ~\cite{underwood2020fraz} can handle irregular responses but often require substantially more evaluations. The central challenge is that the response pattern is unknown before tuning and may vary across datasets, fields, and compressors. Consequently, selecting an appropriate search method becomes a trial-and-error process, while testing multiple methods increases the number of compression evaluations and the overall computational cost. 

This challenge motivates us to investigate whether an LLM agent can serve as an adaptive compression-tuning engine by reasoning over observations from previous trials and selecting subsequent configurations without requiring the response pattern to be known in advance or a specialized search procedure to be predefined.

LLMs provide a promising foundation for such an approach because they can combine reasoning with interactions with external tools. Beyond generating text or code, tool-augmented LLMs can plan multi-step workflows, interpret intermediate observations, and select subsequent actions based on the current task state. Applied to compression tuning, these capabilities allow an agent to iteratively invoke compressors, observe compression and reconstruction quality metrics, reason about the search trajectory, and adapt its next EB selection to the user's objective. This motivates the design of \codename as an LLM-driven agent for adaptive lossy compression tuning.


\section{Methodology}
\label{sec:design}
\subsection{Problem Formulation}

Lossy compression performance depends jointly on the input data, compressor, and selected EB . The same EB can produce substantially different compression ratio (CR) values and reconstruction quality across datasets and compressors. For example, an EB that provides a desired quality level with SZ may produce a different result with ZFP because the two compressors use different representations and error-control mechanisms. Consequently, an EB identified for one compressor or scientific field cannot be assumed to transfer directly to another.

Given an input field, a compressor, and an EB, one compression evaluation produces a set of observable metrics, including CR, PSNR, and SSIM. These metrics become available only after executing the corresponding compression workflow and, when required, decompressing the data and computing the requested quality measurements. We therefore treat the mapping from EB to the resulting compression and quality metrics as an unknown black-box response that can only be observed through actual evaluations. Depending on the compressor and data, this response may be nonlinear or {non-monotonic}.

During compression, users can specify one or more requirements over these metrics. A user request may specify constraints, such as maintaining PSNR or SSIM above a given threshold, and an objective, such as maximizing CR or matching a target value. A candidate configuration is feasible only if it satisfies all hard constraints. Among the feasible configurations, the preferred configuration is the one that best satisfies the user-defined objective.

We formulate this task as a {sequential search} under a finite evaluation budget. After each trial, the evaluated EB and the resulting compression and quality metrics are added to the search history. The next EB is selected using the accumulated observations, the user-defined requirements, and the {remaining evaluation budget}. The objective is to identify a suitable configuration using as few trials as possible.

When multiple compressors are requested, \codename tunes them sequentially. The search history accumulated from previously evaluated compressors, including EBs and observed compression quality metrics, is retained and made available to guide the initialization and search trajectory of subsequent compressors within the same task. After all requested compressors have been evaluated, \codename compares their best feasible configurations according to the user-defined objective and selects the best configuration overall.

\subsection{Overall Design}



\codename uses a single agent to manage the complete tuning trajectory of each compressor. The operations within a trajectory are tightly coupled because each EB selection depends on the observations collected from previous trials. In particular, selecting the next EB requires considering the evaluated EBs, observed compression and quality metrics, active constraints, and remaining evaluation budget. Maintaining these operations within a single agent allows the evolving search state to remain available throughout the select--execute--observe--decide cycle. We therefore do not decompose the tuning trajectory into separate planning, analysis, and candidate-selection agents.

\begin{figure}[!t]
    \centering
    \includegraphics[width=0.99\columnwidth]{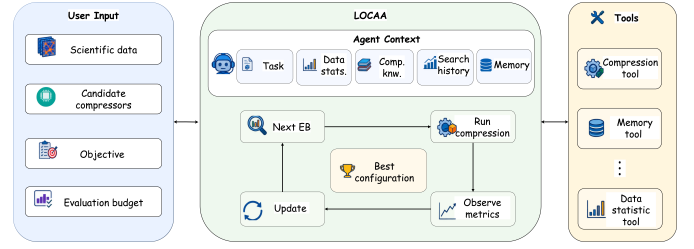}
    \caption{{Overview of the \codename architecture and compression-in-the-loop tuning workflow.}}
    \label{fig:locaa_workflow}
    \vspace{-15pt}
\end{figure}

Fig.~\ref{fig:locaa_workflow} illustrates the overall \codename workflow. The remainder of this section describes its main components, including the task-specific prompt design, persistent memory, and tools interface. In our implementation, GPT-5.5~\cite{openai2026gpt55} serves as the underlying LLM, and the agentic workflow is built using the Pocket Agent framework~\cite{pocket_agent_2024}.

\paragraph{{Task-Specific Prompt Design}}
Compressor tuning can be posed to a general-purpose, tool-augmented language model using a minimal description of compression, decompression, EBs, and output-quality metrics. Although sufficient to define the task, such a description does not specify how to search the EB space efficiently: how to initialize the search, interpret irregular metric responses, allocate a limited trial budget, or decide whether further evaluations are useful. We therefore use a task-specific prompt that represents the tuning procedure as a structured search policy, specifying how the agent selects the initial EB, maintains a trial history, interprets compressor-specific response curves, and projects the next candidate from accumulated observations. It also defines the feasibility conditions, budget-aware stopping rules, and final selection procedure. These instructions preserve the flexibility of the LLM while constraining its decisions to the requirements of compressor tuning.

When a PSNR constraint is available, the agent uses the PSNR definition to estimate an initial EB near the target-quality boundary. Given the data value range ($V_r = V_{\max} - V_{\min}$), PSNR is defined as
\begin{equation}
\mathrm{PSNR} = 20 \cdot \log_{10}\left(\frac{V_r}{\mathrm{RMSE}}\right).
\label{eq:psnr-def}
\end{equation}
The EB ($\epsilon$) directly controls the maximum point-wise error, so ($\mathrm{RMSE} \leq \epsilon$) under absolute-mode compression. Using ($\mathrm{RMSE} = \epsilon$) as the boundary case and inverting Equation~\ref{eq:psnr-def} at the target PSNR gives an initial EB:
\begin{equation}
\epsilon_0 =
\frac{V_r}{10^{\mathrm{PSNR}_{\mathrm{target}}/20}}.
\label{eq:psnr-init}
\end{equation}
This is an approximation rather than an exact feasible bound, since RMSE is generally smaller than $\epsilon$, but it gives a deterministic, dataset-anchored starting point in the correct order-of-magnitude neighborhood of the feasibility boundary, reducing the trials needed to reach a feasible configuration. When no PSNR constraint is specified, the agent randomly selects an initial relative EB from $[10^{-9},\,10^{-1}]$. For compressors operating in absolute-error mode (ABS), the sampled relative EB (REL) is converted using the data value range, $\epsilon_{\mathrm{ABS}} = \epsilon_{\mathrm{REL}} V_r$.

Even without compressor-specific guidance, the agent can apply general heuristics such as increasing the EB to raise CR. These heuristics do not capture every compressor's behavior. ZFP~\cite{Lindstrom2014}, for example, exhibits a staircase-shaped response in which CR and quality metrics change only at discrete bit-plane boundaries, and a range of EBs within a level can produce nearly identical metrics. An agent without this knowledge often selects the next EB by interpolating among previous points, which for ZFP repeatedly places trials within an already observed plateau because no meaningful local slope exists between levels.

We therefore instruct the agent to move between staircase levels using {multiplicative changes} to the EB and to avoid re-evaluating regions that produce previously observed metric values. The search continues until the observed levels bracket the target or the valid EB range is exhausted. If the target lies between two levels, the agent selects the closest feasible observed level and reports that the exact target falls within an unattainable gap. This guidance lets the agent cross plateaus more quickly, avoid redundant evaluations, and reduce the number of compressor calls.

The prompt further requires the agent to follow an explicit observe--interpret--project--act chain-of-thought (CoT)~\cite{wei2022chain} cycle before each compressor evaluation. In the observation stage, the agent examines the EBs and compression quality metrics obtained from previous trials. It then interprets these observations with respect to the tuning objective and uses the accumulated search trajectory to select the next suitable EB. Finally, the agent acts by invoking the compressor with the selected EB, after which the cycle is repeated. By requiring each new decision to account for the accumulated search history, this procedure guides the agent toward more systematic error-bound selection. This design is consistent with prior work showing that explicit state representation and task-specific reasoning procedures can improve the reliability of language-model agents~\cite{shinn2023reflexion,Staller}.



\paragraph{Persistent Memory}
During an active search, \codename evaluates candidate configurations for the current tuning task. After successfully completing the task, \codename stores a compact representation of the search trajectory in persistent memory for reuse in subsequent searches, following prior agent designs that preserve past experience to inform future decisions~\cite{shinn2023reflexion}. The stored record summarizes the completed tuning task, including the evaluated EBs, corresponding compression and quality metrics, agent decisions, and the resulting search strategy. As a representative example, we measured the memory record generated by a five-compressor optimization task that minimized end-to-end time subject to user-defined quality constraints. The complete record occupied 2.1~KB, corresponding to approximately 760 tokens under the OpenAI tokenizer used in our implementation. Thus, a relevant prior record can be provided directly to the agent at the beginning of a new search with limited context overhead.

\codename stores these records in a human-readable text format, allowing the complete search trajectory to be directly inspected and logged. As this work focuses on the agentic workflow design, we use a lightweight storage mechanism that is sufficient for the tuning histories considered in our experiments. Supporting substantially larger tuning histories or cross-dataset knowledge transfer through more scalable storage and retrieval mechanisms is left for future work.

\paragraph{Tools Interface}
The agent interacts with a predefined set of tools exposed through the Model Context Protocol (MCP), an open protocol for connecting language models with external tools and data sources~\cite{anthropic2024mcp}. These tools support compressor execution, dataset characterization, and structured memory access.
\begin{itemize}
    \item \texttt{compression\_runs}: Executes compression and decompression for the specified dataset, dimensions, EB mode, and compressor pool, returning PSNR, SSIM, and CR.
    \item \texttt{get\_initial\_eb}: Returns a random relative EB from \([10^{-9},\,10^{-1}]\); in absolute mode, the agent converts it using the data value range.
    \item \texttt{get\_data\_statistics}: Reports basic statistics of the input, including minimum, maximum, value range, file size, and entropy.
    \item \texttt{write\_memory}: Stores a completed search in a structured format, avoiding the inconsistencies of free-form outputs and enabling reliable reuse.
    \item \texttt{check\_memory}: Queries structured memory for entries matching a given application and data field, returning stored dataset characteristics, configurations, and outcomes to guide the current search.
\end{itemize}
The tool set is extensible: a new compressor is added through \texttt{compression\_runs}, and a new metric by extending the tool outputs. Once exposed through MCP, the agent can use them in subsequent searches.
\section{Evaluation}
\label{sec:eavl}
\subsection{Evaluation Setup}

To assess \codename's performance and validate the design choices described earlier, we conduct a comprehensive empirical evaluation and report the results in this section. Specifically, our evaluation tries to answer the following questions:

\begin{itemize}
\item \textbf{RQ1}: How does \codename perform relative to other search strategies, and how efficiently can it identify EBs that satisfy different user-defined quality constraints?
\item \textbf{RQ2}: Can \codename effectively leverage prior tuning knowledge stored in its memory to identify compression configurations that satisfy new user-defined constraints?
\item \textbf{RQ3}: How do \codename's individual design components contribute to its overall search efficiency?
\end{itemize}
For RQ1, we compare \codename with existing search strategies in terms of constraint satisfaction, achieved objective value, and the number of compression evaluations required. For RQ2, we compare \codename with and without memory initialization to quantify the benefit of reusing prior tuning knowledge. For RQ3, we conduct ablation studies to evaluate the contribution of each key design component. Unless otherwise stated, each LLM-based configuration is executed five independent times, and the average number of trials is reported to account for the stochastic nature of LLM-generated search decisions.

\paragraph{Datasets and Compressors}

We evaluate \codename using multiple scientific data fields drawn from three representative applications: NYX~\cite{nyx2021}, Hurricane Isabel~\cite{isabel2004}, and CESM simulation~\cite{Hurrell2013}. The evaluation considers five error-bounded lossy compressors: SZ~\cite{di2016fast}, SZ3~\cite{SZ3}, SZx~\cite{SZx}, SPERR~\cite{sperr}, and ZFP~\cite{Lindstrom2014}. Because SPERR and ZFP do not natively support the absolute EB mode used by our compression interface, we convert each requested absolute bound (ABS) into an equivalent relative bound (REL) using
$\mathrm{REL}=\mathrm{ABS}/\mathrm{data\ range}$ and execute both compressors in relative EB mode. These datasets and compressors provide diverse data characteristics and compression behaviors, allowing us to evaluate \codename across multiple scientific domains and compressor backends.

\begin{table*}[h]
\centering
\setlength{\tabcolsep}{7pt}
\renewcommand{\arraystretch}{1}
\caption{Number of trials required to achieve target CRs for three compressors on twelve data fields from two applications.}
\label{tab:search-comparison}
\vspace{0pt}

\begin{threeparttable}
\resizebox{0.95\textwidth}{!}{%
\begin{tabular}{ccccccccccccccc}
\hline\hline

\multirow{2}{*}{\textbf{Application}}
& \multirow{2}{*}{\textbf{Fields}}
& \multirow{2}{*}{\textbf{Target}}
&
& \multicolumn{3}{c}{\textbf{SZ}}
&
& \multicolumn{3}{c}{\textbf{SZ3}}
&
& \multicolumn{3}{c}{\textbf{SPERR}}
\\

\cline{5-7}
\cline{9-11}
\cline{13-15}

& & &
& \textbf{BS} & \textbf{FRaZ} & \textbf{LOCAA}
& & \textbf{BS} & \textbf{FRaZ} & \textbf{LOCAA}
& & \textbf{BS} & \textbf{FRaZ} & \textbf{LOCAA}
\\
\hline

\multirow{12}{*}{\textbf{NYX}}
& BD
& \multirow{6}{*}{50}
&
& 8 & 12 & \textbf{5}
& & 7 & 12 & \textbf{3}
& & 8 & 12 & \textbf{2}
\\

& TEMP
&
&
& 8 & 22 & \textbf{5}
& & 7 & 18 & \textbf{4}
& & 7 & 18 & \textbf{1}
\\

& VX
&
&
& 6 & 16 & \textbf{4}
& & 7 & 24 & \textbf{4}
& & 8 & 28 & \textbf{3}
\\

& VY
&
&
& 7 & 18 & \textbf{5}
& & 9 & 24 & \textbf{4}
& & 9 & 24 & \textbf{1}
\\

& VZ
&
&
& 9 & 18 & \textbf{5}
& & 9 & 22 & \textbf{4}
& & 9 & 26 & \textbf{3}
\\

& DMD
&
&
& 6 & 13 & \textbf{5}
& & 9 & 18 & \textbf{3}
& & 8 & 22 & \textbf{3}
\\
\cline{2-15}

& BD
& \multirow{6}{*}{75}
&
& 6 & 20 & \textbf{4}
& & 7 & 18 & \textbf{3}
& & 8 & 16 & \textbf{3}
\\

& TEMP
&
&
& 8 & 16 & \textbf{5}
& & 4 & 24 & \textbf{3}
& & 8 & 20 & \textbf{2}
\\

& VX
&
&
& 9 & 18 & \textbf{5}
& & 9 & 22 & \textbf{4}
& & 9 & 18 & \textbf{2}
\\

& VY
&
&
& 9 & 18 & \textbf{5}
& & 9 & 30 & \textbf{4}
& & 7 & 26 & \textbf{1}
\\

& VZ
&
&
& 8 & 18 & \textbf{6}
& & 8 & 16 & \textbf{4}
& & 9 & 16 & \textbf{3}
\\

& DMD
&
&
& 7 & 22 & \textbf{5}
& & 9 & 22 & \textbf{3}
& & 6 & 26 & \textbf{3}
\\
\hline

\multirow{12}{*}{\textbf{Hurricane}}
& CLOUD
& \multirow{6}{*}{50}
&
& 8 & 17 & \textbf{5}
& & 3 & 25 & \textbf{1}
& & 8 & 27 & \textbf{5}
\\

& P
&
&
& \textbf{4} & 27 & 6
& & \textbf{1} & 13 & 3
& & 8 & 27 & \textbf{2}
\\

& PRECIP
&
&
& 5 & \textbf{3} & 4
& & 8 & 19 & \textbf{3}
& & 6 & 29 & \textbf{3}
\\

& QCLOUD
&
&
& 8 & 5 & \textbf{4}
& & 8 & 11 & \textbf{4}
& & 7 & 19 & \textbf{6}
\\

& TC
&
&
& 9 & 19 & \textbf{6}
& & 9 & 19 & \textbf{3}
& & 9 & 19 & \textbf{3}
\\

& V
&
&
& 9 & 28 & \textbf{6}
& & 8 & 17 & \textbf{4}
& & 8 & 28 & \textbf{3}
\\
\cline{2-15}

& CLOUD
& \multirow{6}{*}{75}
&
& 8 & \textbf{--} & \textbf{6}
& & 8 & 6 & \textbf{3}
& & 8 & 11 & \textbf{4}
\\

& P
&
&
& 7 & 17 & \textbf{5}
& & 8 & \textbf{--} & \textbf{4}
& & 7 & 11 & \textbf{2}
\\

& PRECIP
&
&
& 6 & 31 & \textbf{5}
& & 7 & 19 & \textbf{3}
& & 8 & 19 & \textbf{3}
\\

& QCLOUD
&
&
& 7 & \textbf{3} & 4
& & 7 & 27 & \textbf{3}
& & 8 & 19 & \textbf{5}
\\

& TC
&
&
& \textbf{5} & 19 & \textbf{5}
& & 9 & 23 & \textbf{4}
& & 8 & 19 & \textbf{3}
\\

& V
&
&
& 7 & \textbf{5} & \textbf{5}
& & 5 & 20 & \textbf{4}
& & 6 & 11 & \textbf{3}
\\
\hline

\hline\hline
\end{tabular}%
}

\begin{tablenotes}
\footnotesize
\item[] BS denotes binary search. Bold values indicate the lowest
number of compression evaluations for each compressor.
An entry with "--" indicates that such a configuration cannot be found to
match the target CR within the specified
tolerance.
\end{tablenotes}

\end{threeparttable}
\vspace{-5pt}
\end{table*}

\paragraph{Baseline Search Strategies}

To answer RQ1, we compare \codename with three baseline search strategies selected according to the objective of each use case: binary search (BS), FRaZ~\cite{underwood2020fraz}, and joint binary search (Joint BS). For Use Case~1, which searches for a predefined target CR, we compare \codename with BS and FRaZ. For Use Case~2, which selects the compressor--EB configuration that maximizes CR under joint PSNR and SSIM constraints, we compare \codename with Joint BS. Joint BS performs a constraint-aware EB search for each compressor and then selects the best feasible configuration across the compressor pool. The detailed search procedures, stopping criteria, and rationale for excluding FRaZ from Use Case~2 are presented in the corresponding use-case subsections.

We measure search efficiency using the number of candidate EB evaluations required to identify a feasible configuration. Each evaluated EB is counted as one search evaluation. For tasks involving PSNR or SSIM, an evaluation includes compression, decompression, and quality metric calculation, whereas fixed-ratio FRaZ determines feasibility using the compressed output size. Because the computational work performed during an individual evaluation differs across these search tasks, we use the number of evaluations as the primary measure of search efficiency.

\subsection{RQ1: Benefits of the Agentic Solution}

In practice, compression selection involves diverse objectives, ranging from satisfying a single constraint to optimizing multiple competing objectives. The appropriate objective depends largely on the user's task and performance requirements. To evaluate both the search performance and decision-making capability of the agent, we define three use cases with different constraints and optimization objectives. For detailed analysis, we select three representative examples from each use case.

The three use cases are defined as follows:
\begin{itemize}
    \item \textbf{Use Case 1:} Find an EB that achieves a target CR for a given compressor within a specified tolerance.
    
    \item \textbf{Use Case 2:} Select the compressor--EB pair that maximizes CR subject to
    $\mathrm{PSNR} \geq \mathrm{PSNR}_{\mathrm{target}}$ and
    $\mathrm{SSIM} \geq \mathrm{SSIM}_{\mathrm{target}}$.
    
    \item \textbf{Use Case 3:} Select the compressor--EB pair that minimizes the overall data transfer time while satisfying the specified quality constraints.
\end{itemize}

Across the three use cases, we evaluate the agentic solutions in terms of solution quality and search efficiency. Solution quality is determined by whether the selected configuration satisfies all specified constraints and, when applicable, by its achieved objective value. Search efficiency is measured by the number of evaluation trials required to identify the final solution. Because LLM-based agents can exhibit nondeterministic behavior~\cite{atil2025nondeterminism}, we repeat each experiment five times and report the average, ensuring that the results are not based on a single potentially unrepresentative run.

\begin{table*}[t]
\centering
\setlength{\tabcolsep}{7pt}
\renewcommand{\arraystretch}{1.15}
\caption{{Number of trials required to identify the maximal CR with PSNR $\geq$ 120 dB and SSIM $\geq$ 0.95 among five compressors, on six data fields from the CESM application.}}
\label{tab:bs-locaa-cesm}
\vspace{0pt}

\begin{threeparttable}

\setlength{\aboverulesep}{1.5pt}
\setlength{\belowrulesep}{1.5pt}
\setlength{\cmidrulesep}{1.5pt}

\resizebox{.99\textwidth}{!}{%
\begin{tabular}{ccccccccccccccc}
\toprule

\multirow{2}{*}{\textbf{Data}}
& \multicolumn{7}{c}{\textbf{Joint BS}}
& \multicolumn{7}{c}{\textbf{LOCAA}}
\\

\cmidrule(lr){2-8}
\cmidrule(lr){9-15}

& \textbf{SZ}
& \textbf{SZ3}
& \textbf{SZx}
& \textbf{ZFP}
& \textbf{SPERR}
& \textbf{Total}
& \shortstack{\textbf{Selection}}

& \shortstack{\textbf{SZ}}
& \shortstack{\textbf{SZ3}}
& \shortstack{\textbf{SZx}}
& \shortstack{\textbf{ZFP}}
& \shortstack{\textbf{SPERR}}
& \shortstack{\textbf{Total}}
& \shortstack{\textbf{Selection}}
\\

\midrule

\textbf{LWCF}
& 8
& 8
& 21
& 21
& 6
& 64
& \textbf{SPERR}
& 3
& 3
& 4
& 4
& 3
& 17
& \textbf{SPERR}
\\

\textbf{FLDS}
& 5
& 5
& 21
& 21
& 6
& 58
& \textbf{SPERR}
& 4
& 3
& 4
& 4
& 3
& 18
& \shortstack{\textbf{SZ3 (2)}\\\textbf{SPERR (3)}}
\\

\textbf{FLDSC}
& 5
& 5
& 21
& 21
& 6
& 58
& \textbf{SPERR}
& 4
& 3
& 4
& 4
& 3
& 18
& \textbf{SPERR}
\\

\textbf{CLDMED}
& 8
& 8
& 21
& 21
& 6
& 64
& \textbf{SPERR}
& 3
& 3
& 4
& 4
& 3
& 17
& \textbf{SPERR}
\\

\textbf{AEROD\_v}
& 5
& 8
& 21
& 21
& 6
& 61
& \textbf{SPERR}
& 3
& 3
& 4
& 4
& 3
& 17
& \textbf{SPERR}
\\

\textbf{CLDHGH}
& 8
& 8
& 21
& 21
& 6
& 64
& \textbf{SPERR}
& 3
& 2
& 3
& 4
& 3
& 15
& \textbf{SPERR}
\\

\bottomrule
\end{tabular}%
}

\begin{tablenotes}
\footnotesize
\item Both Joint BS and \codename first identify, for each compressor, the best observed EB that maximizes CR while satisfying the quality constraints and then select the compressor--EB pair with the highest CR. LOCAA's trial counts are averaged across five independent runs. 
{For FLDS, LOCAA selected SZ3 in two runs and SPERR in three runs; all selected configurations satisfied both the PSNR and SSIM constraints.}
\end{tablenotes}

\end{threeparttable}
\vspace{-5pt}
\end{table*}

\paragraph{Use Case 1}
We impose a single compression-ratio constraint and evaluate the search methods on the NYX and HURRICANE applications. For both applications, the target CRs are
$\mathrm{CR}_{\mathrm{target}}\in\{50,75\}$, with a relative tolerance of $\pm2\%$. The corresponding feasible intervals are $[49,51]$ and $[73.5,76.5]$, respectively. All methods use the same datasets, compressor backends, relative EB range of $[10^{-9},10^{-1}]$, feasibility criterion, and maximum budget of 60 compression evaluations. A search is successful only if the achieved CR falls within the corresponding interval; otherwise, it is denoted by ``--'' in Table~\ref{tab:search-comparison}.

Binary search can identify a feasible configuration efficiently when CR varies approximately monotonically with the relative EB, but its effectiveness may decrease for irregular or locally non-monotonic responses. FRaZ instead treats the compressor as a black box and performs a broader global search, improving robustness at the cost of more trials.

\codename is evaluated using five independent runs per configuration. Memory is cleared between runs so that each run begins from the same initial state. 
Because SZx and ZFP exhibit staircase-shaped compression-ratio responses for which a specified target may be unattainable, they are excluded from this experiment. Table~\ref{tab:search-comparison} therefore reports results for SZ, SZ3, and SPERR.

Across all fields, targets, and compressors, \codename requires an average of 3.76 trials, compared with 7.44 for binary search and 18.93 for FRaZ, corresponding to approximately 1.98$\times$ and 5.03$\times$ reduction, respectively. For the NYX application, \codename requires approximately 2.18$\times$ fewer trials than binary search and 5.53$\times$ fewer than FRaZ. For HURRICANE, it requires approximately 1.80$\times$ and 4.56$\times$ fewer trials, respectively. In two cases, FRaZ fails to reach the target within the budget.

\paragraph{Use Case 2}
This use case extends single-compressor tuning to joint compressor and EB selection. We evaluate \codename on the CESM application, where it sequentially searches the compressor pool and selects the compressor--EB pair with the highest CR subject to $\mathrm{PSNR}\geq120$~dB and $\mathrm{SSIM}\geq0.95$. FRaZ is excluded because it is designed for fixed-ratio search, which is not applicable here.

We compare \codename with a modified binary-search baseline, referred to as Joint BS, under identical experimental settings. Joint BS is designed specifically to select a single EB that jointly satisfies the PSNR and SSIM constraints. For each compressor, it evaluates the geometric midpoint of the current EB interval and checks whether both constraints are satisfied. If they are, Joint BS updates the lower bound to explore larger EBs and potentially achieve a higher CR; otherwise, it updates the upper bound to preserve reconstruction quality. This update rule assumes that PSNR and SSIM change monotonically with the EB. Therefore, Joint BS may fail or converge to a suboptimal configuration when either metric exhibits a non-monotonic or irregular response, because satisfying or violating both constraints at one EB does not necessarily determine the appropriate search direction.
The search terminates upon reaching the interval limit, quality tolerance
($\mathrm{TOL}_{\mathrm{PSNR}}=1$~dB or
$\mathrm{TOL}_{\mathrm{SSIM}}=0.005$), or evaluation budget.

\begin{figure}
    \centering
    \vspace{-15pt}
    \includegraphics[width=0.85\linewidth]{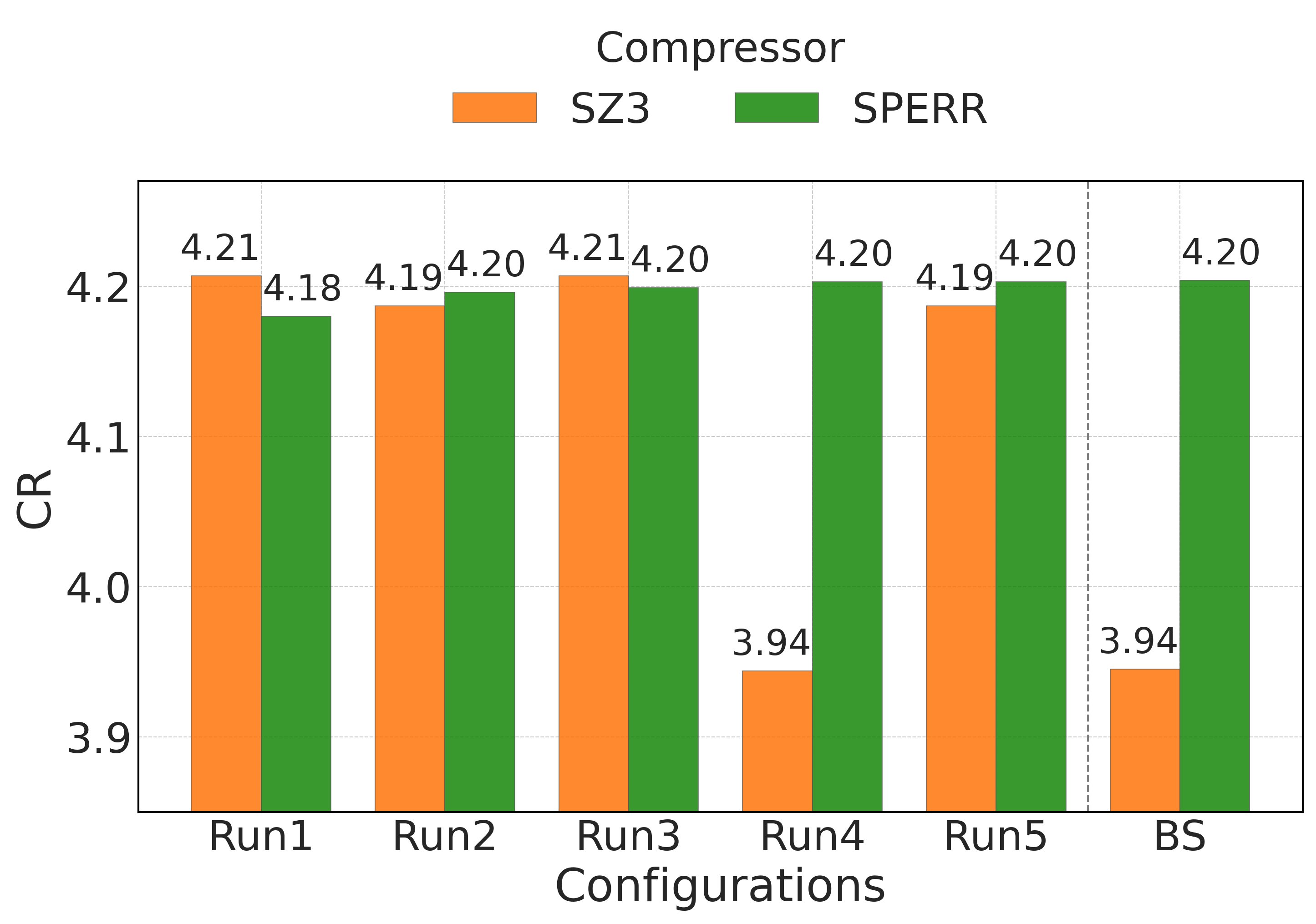}
    \caption{{Best feasible CR for SZ3 and SPERR on CESM FLDS under PSNR $\geq 120$~dB and SSIM $\geq 0.95$ across five \codename runs and the BS baseline.}}
    \label{fig:Runs_variation}
    \vspace{-15pt}
\end{figure}

Table~\ref{tab:bs-locaa-cesm} shows that \codename requires 15--18 trials per CESM field, compared with 58--64 for Joint BS, with an average reduction of 72.4\%. Joint BS requires particularly more trials for SZx and ZFP because of their staircase-shaped responses. Joint BS selects SPERR for all six fields, while \codename selects SPERR for five. For FLDS, \codename selects SPERR in three runs and SZ3 in two, as shown in Fig.~\ref{fig:Runs_variation}. This variation in compressor selection occurs in Runs~1 and~3 because \codename does not converge as closely to the quality-constraint boundary as Joint BS, resulting in slightly different feasible CRs across compressors. Both selected configurations satisfy the PSNR and SSIM constraints. Because all selected configurations satisfy both quality constraints, we compare the search efficiency of the two approaches using the number of trials required to obtain the final solution.

\begin{table*}[t!]
\centering
\caption{{Best-fit compressor configurations to minimize the total transfer time across a 100 Mbps network. Compressor time includes compression, decompression, and quality metric calculation; the selected configuration minimizes total compressor and transfer time.}}
\label{tab:compressor_selection}
\resizebox{.99\textwidth}{!}{%
\begin{tabular}{lllcrrrrrc}
\toprule
\multirow{2}{*}{\textbf{Application}} &
\multirow{2}{*}{\textbf{Data}} &
\multirow{2}{*}{\textbf{Size (MiB)}} &
\multirow{2}{*}{\textbf{Compressor}} &
\multirow{2}{*}{\textbf{REL EB}} &
\multirow{2}{*}{\textbf{CR}} &
\multicolumn{3}{c}{\textbf{Time (s)}} &
\multirow{2}{*}{\textbf{Selected}} \\
\cmidrule(lr){7-9}
& & & & & &
\textbf{Compressor Time} &
\textbf{I/O Transfer} &
$\boldsymbol{T_{\mathrm{total}}}$ & \\
\midrule

\multirow{5}{*}{NYX}
& \multirow{5}{*}{BD}
& \multirow{5}{*}{512.00}
& SZ
& $6.500\times10^{-7}$
& 28.345
& 7.544113
& 1.515247
& 9.059360
& \\

& & & SZ3
& $6.500\times10^{-7}$
& 53.924
& 9.187456
& 0.796485
& 9.983941
& \\

& & & SZx
& $1.450\times10^{-6}$
& 4.359
& 3.179106
& 9.853102
& 13.032208
& \\

& & & ZFP
& $1.200\times10^{-5}$
& 30.600
& 6.268689
& 1.403584
& 7.672273
& \\

& & & \textbf{SPERR}
& $\mathbf{2.400\times10^{-6}}$
& \textbf{121.277}
& \textbf{7.309494}
& \textbf{0.354145}
& \textbf{7.663639}
& $\checkmark$ \\

\midrule

\multirow{5}{*}{CESM}
& \multirow{5}{*}{CLDMED}
& \multirow{5}{*}{24.72}
& SZ
& $1.700\times10^{-6}$
& 3.177
& 0.672265
& 0.652691
& 1.324956
& \\

& & & SZ3
& $1.700\times10^{-6}$
& 3.361
& 1.660266
& 0.616959
& 2.277225
& \\

& & & \textbf{SZx}
& $\mathbf{7.000\times10^{-6}}$
& \textbf{1.822}
& \textbf{0.136440}
& \textbf{1.138090}
& \textbf{1.274530}
& $\checkmark$ \\

& & & ZFP
& $7.000\times10^{-6}$
& 2.420
& 0.763383
& 0.856860
& 1.620243
& \\

& & & SPERR
& $1.700\times10^{-6}$
& 3.245
& 3.790395
& 0.639014
& 4.429409
& \\

\midrule

\multirow{5}{*}{HURRICANE}
& \multirow{5}{*}{CLOUD}
& \multirow{5}{*}{95.37}
& SZ
& $5.000\times10^{-6}$
& 23.441
& 1.326248
& 0.341282
& 1.667530
& \\

& & & SZ3
& $4.000\times10^{-6}$
& 19.110
& 1.799290
& 0.418629
& 2.217919
& \\

& & & \textbf{SZx}
& $\mathbf{7.000\times10^{-6}}$
& \textbf{8.251}
& \textbf{0.325019}
& \textbf{0.969579}
& \textbf{1.294598}
& $\checkmark$ \\

& & & ZFP
& $2.000\times10^{-5}$
& 12.100
& 0.932791
& 0.661157
& 1.593948
& \\

& & & SPERR
& $3.500\times10^{-6}$
& 8.353
& 3.487780
& 0.957740
& 4.445520
& \\

\bottomrule
\end{tabular}%
}
\end{table*}

\paragraph{Use Case 3}
We construct a more challenging compressor selection task that jointly considers CR and compression--decompression time. Given reconstruction quality constraints and a specified I/O bandwidth, the agent must select the compressor and EB configuration that minimizes the total end-to-end time. Maximizing the CR alone may produce a suboptimal configuration because the resulting reduction in I/O time can be offset by high compression and decompression overhead. The optimal configuration must therefore balance compression efficiency against the computational cost of compression and reconstruction.
For this experiment, we require $\mathrm{SSIM} \geq 0.95$ and $\mathrm{PSNR} \geq 120$~dB and assume a nominal network bandwidth of 100 Mbps, representing wide-area data transfers between laboratory and university sites.

Given an original data size $S_{\mathrm{original}}$ in MiB and compression
ratio $\mathrm{CR}$, the compressed data size is
$S_{\mathrm{original}}/\mathrm{CR}$ MiB. For a network bandwidth
$B_{\mathrm{net}}$ measured in bits per second, the compressed-data transfer
time is estimated as
\begin{equation}
T_{\mathrm{transfer}} =
\frac{8 \times 2^{20} \times S_{\mathrm{original}}}
{\mathrm{CR} \times B_{\mathrm{net}}}.
\label{eq:compressed_transfer_time}
\end{equation}
For this experiment, compressor time is defined as
\begin{equation}
T_{\mathrm{compressor}} =
T_{\mathrm{comp}} +
T_{\mathrm{decomp}} +
T_{\mathrm{metric}},
\label{eq:compressor_time}
\end{equation}
where $T_{\mathrm{metric}}$ denotes the time required to compute the
reconstruction quality metrics, including SSIM and PSNR. The total objective
time is then
\begin{equation}
T_{\mathrm{total}} =
T_{\mathrm{compressor}} + T_{\mathrm{transfer}}.
\label{eq:total_data_movement_time}
\end{equation}

Table~\ref{tab:compressor_selection} compares compressor configurations that satisfy the reconstruction quality constraints. Because compressor time (compression, decompression, and quality metric computation) can offset the transfer time saved by a higher CR, the final selection depends jointly on CR, data size, network bandwidth, and compressor time. We evaluated one field from each application, with data sizes from 24.72 to 512.00~MiB. At 100 Mbps, \codename selects SZx for CESM and HURRICANE and SPERR for NYX. For CESM and HURRICANE, SZx yields the lowest total time despite not achieving the highest CR; for NYX, SPERR narrowly outperforms ZFP.

These results demonstrate two capabilities of \codename. First, rather than maximizing CR, it identifies the best feasible configuration by balancing CR, compressor time, data size, and bandwidth-dependent transfer time, producing different selections across applications. Second, \codename derives the transfer-time calculation from the natural-language objective even though the corresponding equations are not given in the prompt, translating a user-defined objective into an application-specific selection decision.

\begin{figure}[t]
\centering

\begin{subfigure}[t]{0.49\columnwidth}
    \centering
    \includegraphics[width=\linewidth]{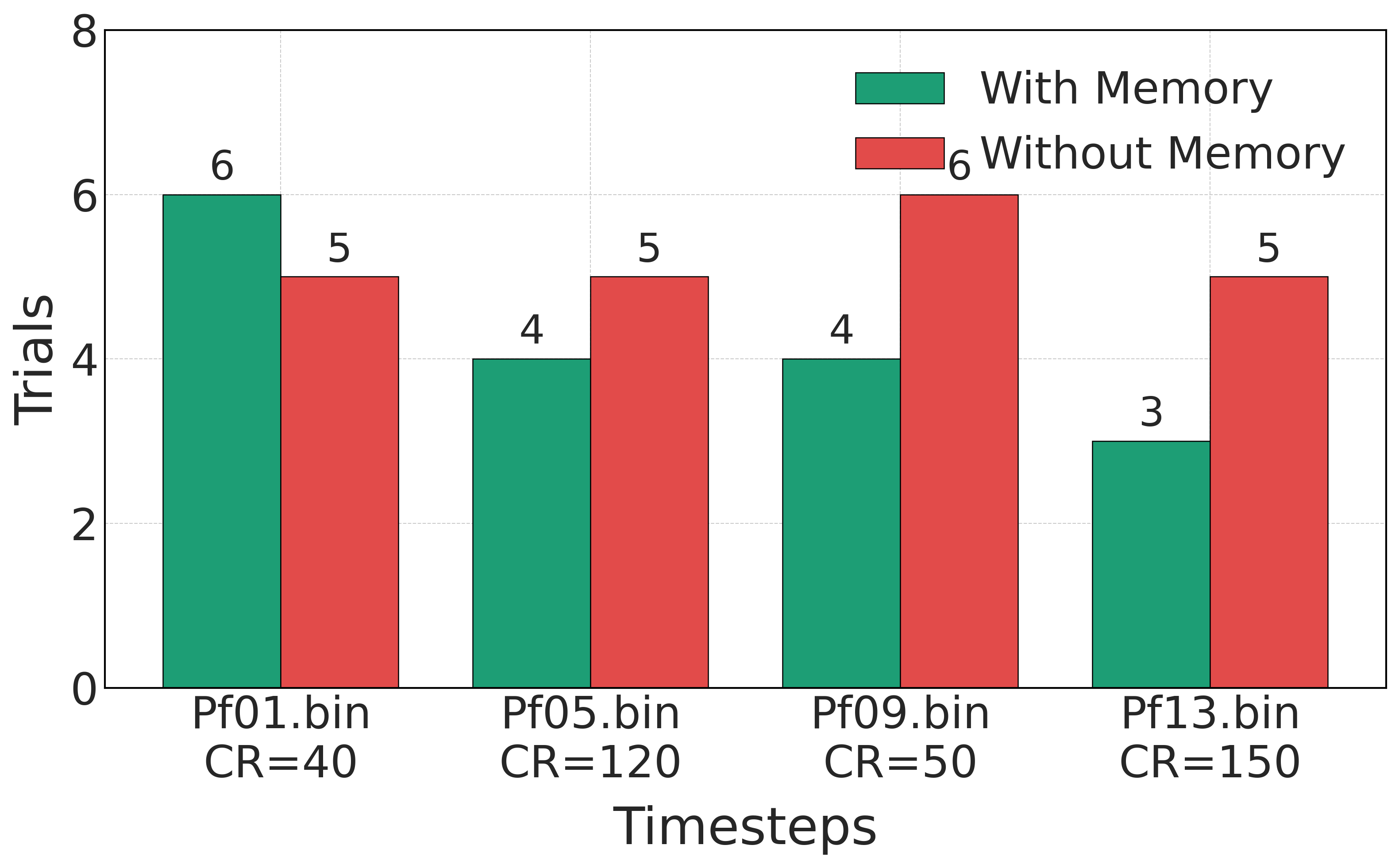}
    \caption{P field.}
    \label{fig:memory_trials_p}
\end{subfigure}
\hfill
\begin{subfigure}[t]{0.49\columnwidth}
    \centering
    \includegraphics[width=\linewidth]{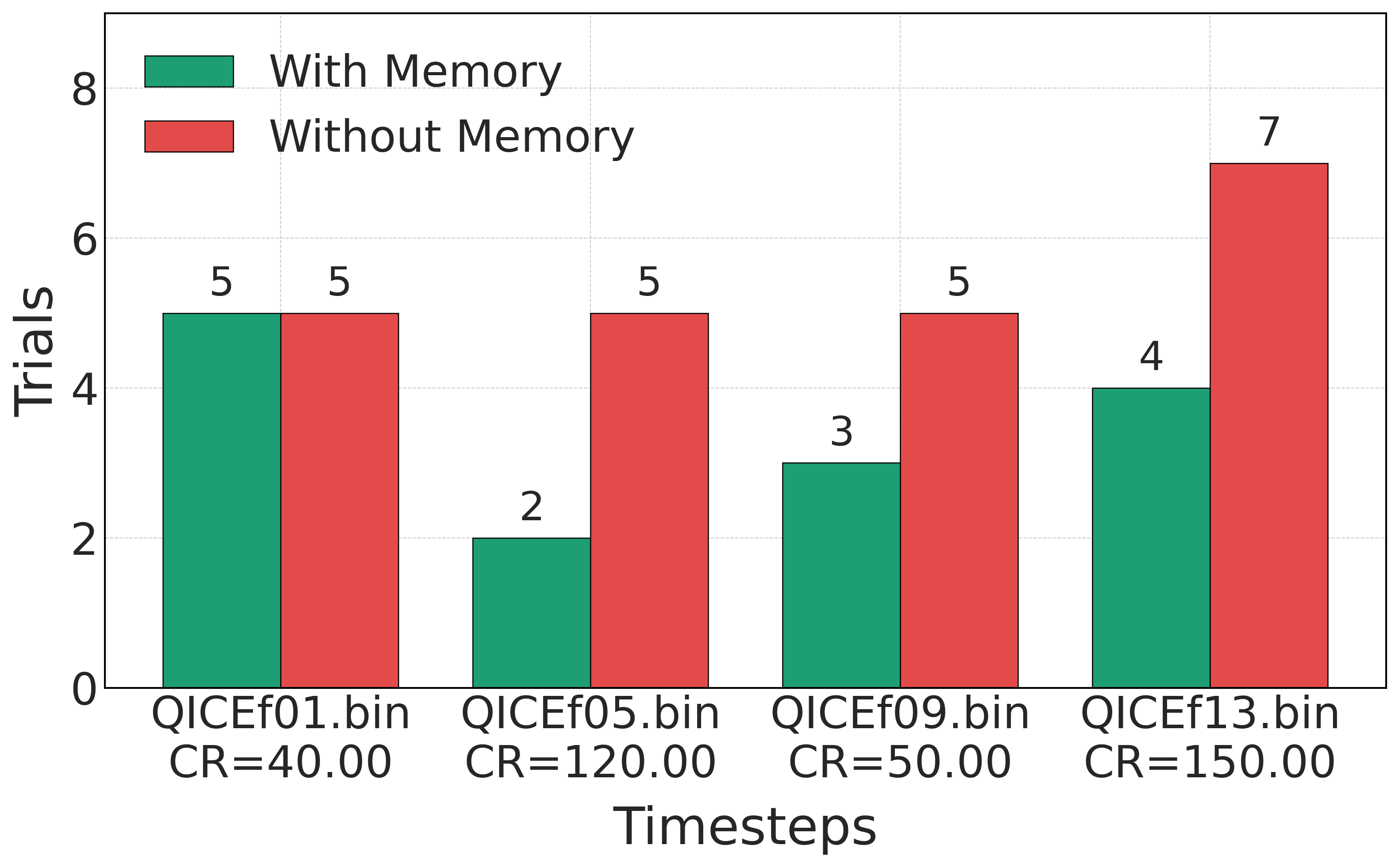}
    \caption{QICE field.}
    \label{fig:memory_trials_qice}
\end{subfigure}

\caption{{Tuning efficiency of \codename with and without persistent memory. Datasets: P and QICE; Compressor: SZ3.}}
\label{fig:memory_trials}

\vspace{-8pt}
\end{figure}

\begin{table*}[t]
\centering
\caption{{Ablation of four configurations combining CoT and compressor-specific knowledge across six CESM fields for maximizing CR while maintaining the quality constraints. Lower is better.}}
\label{tab:cot_compression_knowledge}
\renewcommand{\arraystretch}{1.15}
\setlength{\tabcolsep}{3pt}

\resizebox{.99\textwidth}{!}{%
\begin{tabular}{ll*{20}{c}}
\toprule
& &
\multicolumn{5}{c}{\makecell{\textbf{Without CoT, without Comp. Knw.}}} &
\multicolumn{5}{c}{\makecell{\textbf{With CoT, without Comp. Knw.}}} &
\multicolumn{5}{c}{\makecell{\textbf{Without CoT, with Comp. Knw.}}} &
\multicolumn{5}{c}{\makecell{\textbf{With CoT, with Comp. Knw.}}} \\
\cmidrule(lr){3-7}
\cmidrule(lr){8-12}
\cmidrule(lr){13-17}
\cmidrule(lr){18-22}

\textbf{Data} &
\textbf{Target} &
\textbf{SZ} & \textbf{SZ3} & \textbf{SZx} & \textbf{ZFP} & \textbf{SPERR} &
\textbf{SZ} & \textbf{SZ3} & \textbf{SZx} & \textbf{ZFP} & \textbf{SPERR} &
\textbf{SZ} & \textbf{SZ3} & \textbf{SZx} & \textbf{ZFP} & \textbf{SPERR} &
\textbf{SZ} & \textbf{SZ3} & \textbf{SZx} & \textbf{ZFP} & \textbf{SPERR} \\
\midrule

AEROD\_v &
\multirow{6}{*}{\makecell{SSIM $\geq 0.95$\\
PSNR $\geq 97$ dB}} &
4 & 4 & 10 & 8 & 5 &
4 & 5 & 9 & 7 & 4 &
4 & 4 & 8 & 4 & 5 &
3 & 6 & 10 & 5 & 5 \\

CLDHGH & &
3 & 4 & 10 & 9 & 4 &
2 & 3 & 10 & 8 & 3 &
3 & 3 & 9 & 4 & 5 &
3 & 4 & 8 & 4 & 4 \\

CLDMED & &
4 & 4 & 6 & 9 & 4 &
3 & 4 & 4 & 6 & 3 &
3 & 4 & 6 & 4 & 3 &
4 & 4 & 6 & 4 & 4 \\

FLDS & &
4 & 4 & 9 & 9 & 4 &
3 & 4 & 8 & 7 & 5 &
3 & 3 & 10 & 4 & 4 &
3 & 5 & 8 & 4 & 4 \\

FLDSC & &
3 & 5 & 10 & 9 & 4 &
3 & 5 & 8 & 6 & 4 &
5 & 5 & 9 & 4 & 4 &
3 & 5 & 8 & 6 & 5 \\

LWCF & &
3 & 4 & 8 & 8 & 4 &
3 & 4 & 7 & 8 & 3 &
3 & 4 & 8 & 4 & 4 &
3 & 3 & 10 & 4 & 4 \\

\bottomrule
\end{tabular}%
}
\end{table*}

\subsection{RQ2: Memory Evaluation}

\codename's memory preserves previous tuning outcomes, including evaluated compressors, EBs, and compression quality metrics. To answer RQ2, we examine whether this knowledge can guide subsequent searches and reduce compression--decompression evaluations. For this experiment, we evaluate memory reuse across different timesteps of the same physical field rather than across different data fields.

We compare memory-enabled and memory-free \codename configurations using SZ3 on the HURRICANE pressure (P) and cloud-ice (QICE) fields. Each field includes four timesteps, \texttt{f01}, \texttt{f05}, \texttt{f09}, and \texttt{f13}, with target CRs of 40, 120, 50, and 150, respectively. The configurations use identical EB ranges, constraints, and evaluation budgets and differ only in whether prior results from the same physical field are available at initialization. Search efficiency is measured by the number of evaluations required to find a feasible configuration. Each configuration is executed five independent times, and we report the mean number of evaluations required to identify a feasible configuration.

 Fig.~\ref{fig:memory_trials} shows that memory reduces the number of trials in six of the eight cases. For the P field, the trial counts for Pf05, Pf09, and Pf13 decrease from 5 to 4, 6 to 4, and 5 to 3, respectively, while the count for Pf01 increases from 5 to 6. Consequently, the average decreases from 5.25 to 4.25 trials, representing a 19.0\% reduction. For QICE, memory matches the five-trial baseline for QICEf01 and reduces the trial counts for the remaining three timesteps from 5 to 2, 5 to 3, and 7 to 4. This decreases the QICE average from 5.50 to 3.50 trials, representing a 36.4\% reduction. Across both fields, memory reduces the overall average from 5.38 to 3.88 trials, corresponding to a 27.9\% reduction. These results indicate that prior tuning knowledge can improve search efficiency across timesteps of the same field, although its benefit depends on the relevance of the stored information to the new task.
 
We did not evaluate cross-field memory transfer. Basic summary statistics such as value range, standard deviation, and entropy may be insufficient to identify similar error-bound--metric responses, which also depend on spatial structure, compressor characteristics, and the selected EB~\cite{underwood2023black,krasowska2021exploring}. Identifying reliable indicators for cross-field knowledge transfer is left for future work.


\subsection{Ablation Study}
To answer RQ3, we perform a series of ablation experiments to quantify the contribution of \codename's key design components. We first jointly evaluate the {effects of CoT} and {compressor-specific knowledge} using different combinations of the two components. We then separately evaluate the effect of PSNR-driven initialization on \codename's search decisions. Across these experiments, the remaining settings are kept unchanged to isolate the contribution of the component under study. We examine their effects primarily in terms of search efficiency and the resulting compression configuration.
\paragraph{Effect of CoT and compressor-specific knowledge}
We evaluate two components of \codename's task-specific prompting strategy: CoT guidance, which provides explicit per-trial reasoning for interpreting each evaluation and selecting the next EB, and compressor-specific knowledge, which describes the characteristic behavior of individual compressors. We test all four combinations of the two components on six CESM fields with PSNR- and SSIM-based quality targets, maximizing CR for each compressor. PSNR-driven initialization is disabled throughout to isolate these two components.

Table~\ref{tab:cot_compression_knowledge} enables each contribution to be examined by changing one component at a time. Adding CoT alone (\textit{[with CoT, without Comp. Knw.]} vs. \textit[without CoT, without Comp. Knw.]) reduces the overall average from 5.87 to 5.10 trials, a 13.1\% reduction, with per-compressor gains for SZ, SZx, ZFP, and SPERR and no change for SZ3. Compressor knowledge alone (\textit{[without CoT, with Comp. Knw.]} vs.\ [without CoT, without Comp. Knw.]) reduces the average from 5.87 to 4.77 trials, an 18.8\% reduction, driven almost entirely by ZFP, which falls from 8.67 to 4.00 trials (53.8\%, a $2.17\times$ speedup); other compressors change little. This indicates that CoT refines the generic search, whereas compressor-specific knowledge is decisive when the response departs from a smooth EB curve, as with ZFP's staircase behavior. The combined \textit{[with CoT, with Comp. Knw.]} configuration requires 4.97 trials on average, a 15.3\% reduction over \textit[without CoT, without Comp. Knw.] but slightly more than knowledge alone (4.77). Adding CoT on top of knowledge helps SZ but raises the averages for SZ3, ZFP, and SPERR, so the two components are not strictly additive.

\begin{figure}
    \centering
    \vspace{-15pt}
    \includegraphics[width=0.8\linewidth]{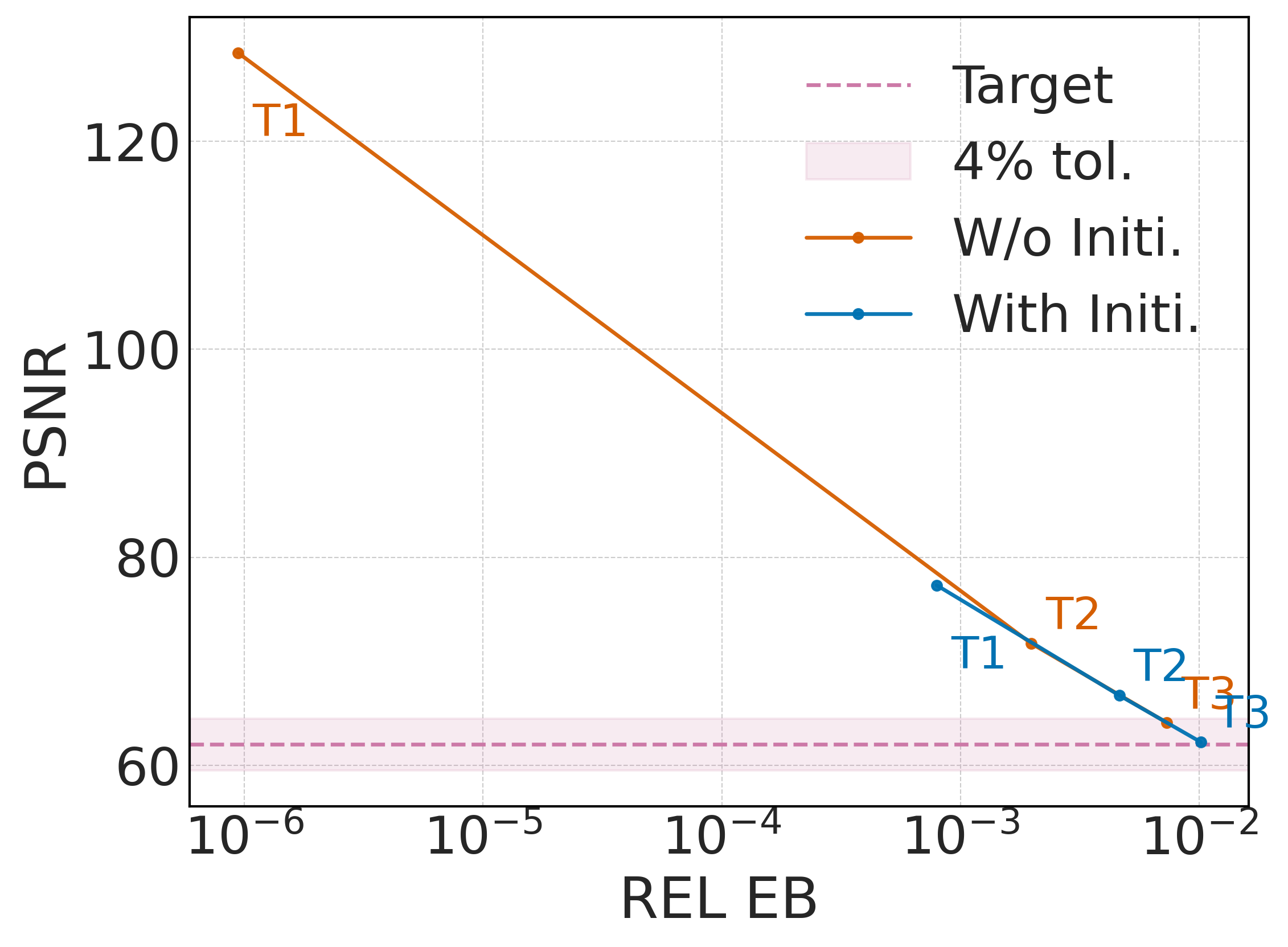}
    \caption{Both \codename configurations reach PSNR $62\pm4\%$ in three trials on CESM FLDSC with SPERR.}
    \label{fig:sperr_trials}
\end{figure}

\paragraph{Initialization}
 \codename uses the target PSNR to estimate an initial EB near the desired quality level. We evaluate this strategy on two CESM fields using PSNR targets of 62~dB and 82~dB, a 4\% tolerance, and five compressors. The initialization-enabled agent starts from the PSNR-driven error-bound estimate in Equation~\ref{eq:psnr-init}. The agent without this knowledge starts from a random EB. All other prompt components remain identical, and compressor-specific knowledge is removed from both prompts to isolate the effect of initialization. Because the estimate is approximate and LLM search trajectories can vary across runs~\cite{atil2025nondeterminism}, starting closer to the target does not always reduce the number of trials. Fig.~\ref{fig:sperr_trials} illustrates a case in which both configurations converge in the same number of trials.

Table~\ref{tab:initialization_comparison} shows that PSNR-driven initialization requires fewer trials in 10 of the 20 configurations and matches the performance of the agent without initialization knowledge in the remaining 10, without increasing the trial count in any case. Overall, the total number of trials decreases from 54 to 44, while the average decreases from 2.70 to 2.20 trials per configuration, representing an 18.5\% reduction. Thus, PSNR-driven initialization improves aggregate search efficiency even when it provides no benefit for some individual configurations.
\begin{table}[t]
\centering
\caption{Number of trials with and without EB initialization. Strictly lower
values for each compressor are shown in bold.}
\label{tab:initialization_comparison}
\setlength{\tabcolsep}{3pt}
\renewcommand{\arraystretch}{1.05}
\resizebox{\columnwidth}{!}{%
\begin{tabular}{@{}llcccccccccc@{}}
\toprule
\multirow{2}{*}{\textbf{Data}} &
\multirow{2}{*}{\textbf{Target}} &
\multicolumn{5}{c}{\textbf{W/o Initialization}} &
\multicolumn{5}{c}{\textbf{With Initialization}} \\
\cmidrule(lr){3-7}
\cmidrule(lr){8-12}
& &
\textbf{SZ} & \textbf{SZ3} & \textbf{SZx} & \textbf{ZFP} & \textbf{SPERR} &
\textbf{SZ} & \textbf{SZ3} & \textbf{SZx} & \textbf{ZFP} & \textbf{SPERR} \\
\midrule

CLDHGH &
\multirow{2}{*}{62} &
2 & 3 & 3 & 2 & 2 &
2 & \textbf{2} & \textbf{2} & 2 & 2 \\

FLDSC &
&
3 & 3 & 3 & 4 & 3 &
\textbf{2} & \textbf{2} & 3 & 4 & 3 \\

\midrule

CLDHGH &
\multirow{2}{*}{82} &
2 & 3 & 3 & 3 & 2 &
2 & \textbf{2} & \textbf{2} & \textbf{2} & 2 \\

FLDSC &
&
2 & 2 & 3 & 3 & 3 &
2 & 2 & \textbf{2} & \textbf{2} & \textbf{2} \\

\bottomrule
\end{tabular}%
}
\vspace{-5pt}
\end{table}

\subsection{{Token-Usage Analysis}}
The cost and latency of \codename depend on the number of LLM interactions and
compression evaluations performed during tuning. We measured token usage
on three tuning tasks using the NYX, HURRICANE, and
CESM applications. Each run evaluated five candidate compressors under the
specified SSIM and PSNR constraints to maximize CR.

Across the three evaluated fields, one from each application, a complete
\codename tuning run required an average of 22.3 LLM interactions. Each
interaction processed approximately 9.2K input tokens and generated 0.47K
output tokens, with approximately 8.1K input tokens served from the prompt
cache. Overall, each tuning run used approximately 215.9K tokens, including
205.4K input and 10.5K output tokens. Approximately 181.5K input tokens were
cached, representing 88.4\% of the input-token volume. Our implementation uses the GPT-5.5 API~\cite{openai2026gpt55}, and the associated monetary cost may change over time as inference costs decrease. 

\section{Limitations and Future Work}
\label{sec:limitation}
\codename relies on compression-in-the-loop evaluation: each candidate must be compressed and decompressed before it can be assessed. This gives direct measurements of CR, reconstruction quality, and runtime, but repeated executions add computational overhead for large datasets or expensive compressors, so the tuning cost depends on both the number of trials and the per-workflow execution cost. \codename also incurs the inference cost and latency of LLM-based reasoning.

Our implementation uses only the GPT API, and we have not tested whether the behavior generalizes to other proprietary or open-source models. The evaluation is also at a relatively small scale, with a limited set of datasets, compressors, and constraints. Further work is needed to establish \codename's robustness across larger datasets, broader compressor collections, more complex constraints, and production-scale HPC environments.

\section{Conclusion}
\label{sec:conclusion}
In this study, we proposed \codename, an autonomous tuning agent for scientific lossy compression. By combining agentic large language models with compression-in-the-loop evaluation, \codename efficiently searches compressor and EB configurations while supporting different user-defined compression, quality, and performance objectives. Our evaluation demonstrates that \codename can reason over observed compression outcomes, adapt its search strategy, and select configurations that satisfy reconstruction quality constraints while optimizing the specified objective. This capability distinguishes \codename from conventional tuning approaches that rely on fixed search procedures, compressor-specific models, or extensive configuration evaluations. We believe \codename establishes a promising direction for autonomous and flexible lossy compression tuning in HPC workflows. By allowing users to express application-specific requirements without manually designing optimization algorithms, \codename can make compression tuning more accessible to domain scientists and support more efficient scientific data management. In future work, we plan to evaluate \codename on larger-scale datasets and production HPC systems, extend its support to additional compressors and optimization objectives, and investigate methods for further reducing the number and cost of compression evaluations.

\bibliographystyle{IEEEtran}
\bibliography{main}

@article{atil2025nondeterminism,
  title   = {Non-Determinism of ``Deterministic'' LLM Settings},
  author  = {Atil, Berk and Aykent, Sarp and Chittams, Alexa and Fu, Lisheng
             and Passonneau, Rebecca J. and Radcliffe, Evan
             and Rajagopal, Guru Rajan and Sloan, Adam and Tudrej, Tomasz
             and Ture, Ferhan and Wu, Zhe and Xu, Lixinyu
             and Baldwin, Breck},
  journal = {arXiv preprint arXiv:2408.04667},
  year    = {2025}
}

@inproceedings{wei2022chain,
  title     = {Chain-of-Thought Prompting Elicits Reasoning in Large Language Models},
  author    = {Wei, Jason and Wang, Xuezhi and Schuurmans, Dale and Bosma, Maarten
               and Ichter, Brian and Xia, Fei and Chi, Ed H. and Le, Quoc V.
               and Zhou, Denny},
  booktitle = {Advances in Neural Information Processing Systems},
  volume    = {35},
  pages     = {24824--24837},
  year      = {2022}
}

@INPROCEEDINGS{di2016fast,
  author    = {Di, Sheng and Cappello, Franck},
  booktitle = {2016 IEEE International Parallel and Distributed Processing Symposium (IPDPS)},
  title     = {Fast Error-Bounded Lossy HPC Data Compression with SZ},
  year      = {2016},
  pages     = {730--739},
  doi       = {10.1109/IPDPS.2016.11}
}

@ARTICLE{SZ3,
  author  = {Liang, Xin and Zhao, Kai and Di, Sheng and Li, Sihuan
             and Underwood, Robert and Gok, Ali M. and Tian, Jiannan
             and Deng, Junjing and Calhoun, Jon C. and Tao, Dingwen
             and Chen, Zizhong and Cappello, Franck},
  journal = {IEEE Transactions on Big Data},
  title   = {SZ3: A Modular Framework for Composing Prediction-Based
             Error-Bounded Lossy Compressors},
  year    = {2023},
  volume  = {9},
  number  = {2},
  pages   = {485--498},
  doi     = {10.1109/TBDATA.2022.3201176}
}

@inproceedings{yao2023react,
  title     = {{ReAct}: Synergizing Reasoning and Acting in Language Models},
  author    = {Yao, Shunyu and Zhao, Jeffrey and Yu, Dian and Du, Nan
               and Shafran, Izhak and Narasimhan, Karthik R.
               and Cao, Yuan},
  booktitle = {International Conference on Learning Representations},
  year      = {2023}
}

@article{wang2024llmagents,
  title   = {A Survey on Large Language Model Based Autonomous Agents},
  author  = {Wang, Lei and Ma, Chen and Feng, Xueyang and Zhang, Zeyu
             and Yang, Hao and Zhang, Jingsen and Chen, Zhiyuan
             and Tang, Jiakai and Chen, Xu and Lin, Yankai
             and Zhao, Wayne Xin and Wei, Zhewei and Wen, Ji-Rong},
  journal = {Frontiers of Computer Science},
  volume  = {18},
  number  = {6},
  pages   = {186345},
  year    = {2024},
  doi     = {10.1007/s11704-024-40231-1}
}

@article{kon2025curie,
  title   = {Curie: Toward Rigorous and Automated Scientific
             Experimentation with {AI} Agents},
  author  = {Kon, Patrick Tser Jern and Liu, Jiachen and Ding, Qiuyi
             and Qiu, Yiming and Yang, Zhenning and Huang, Yibo
             and Srinivasa, Jayanth and Lee, Myungjin
             and Chowdhury, Mosharaf and Chen, Ang},
  journal = {arXiv preprint arXiv:2502.16069},
  year    = {2025},
  doi     = {10.48550/arXiv.2502.16069}
}

@misc{nyx2021,
  author       = {{Nyx Development Team}},
  title        = {{Nyx}: An Adaptive Mesh, Cosmological Hydrodynamics
                  Simulation Code},
  year         = {2021},
  howpublished = {\url{https://amrex-astro.github.io/Nyx/}},
  note         = {Online; accessed: 2025-04-02}
}

@misc{openai2026gpt55,
  author       = {{OpenAI}},
  title        = {Introducing {GPT-5.5}},
  year         = {2026},
  month        = apr,
  howpublished = {\url{https://openai.com/index/introducing-gpt-5-5/}},
  note         = {Accessed: 2026-08-06}
}

@article{wang2004ssim,
  author  = {Wang, Zhou and Bovik, Alan C. and Sheikh, Hamid R.
             and Simoncelli, Eero P.},
  title   = {Image Quality Assessment: From Error Visibility
             to Structural Similarity},
  journal = {IEEE Transactions on Image Processing},
  volume  = {13},
  number  = {4},
  pages   = {600--612},
  year    = {2004},
  month   = apr,
  doi     = {10.1109/TIP.2003.819861}
}

@article{Hurrell2013,
  author = {James W. Hurrell and M. M. Holland and P. R. Gent and S. Ghan and Jennifer E. Kay and P. J. Kushner and J. F. Lamarque and W. G. Large and D. Lawrence and K. Lindsay and W. H. Lipscomb and M. C. Long and N. Mahowald and D. R. Marsh and R. B. Neale and P. Rasch and S. Vavrus and M. Vertenstein and D. Bader and W. D. Collins and J. J. Hack and J. Kiehl and S. Marshall},
  doi = {10.1175/BAMS-D-12-00121.1},
  issn = {00030007},
  issue = {9},
  journal = {Bulletin of the American Meteorological Society},
  month = {9},
  pages = {1339-1360},
  publisher = {American Meteorological Society},
  title = {The Community Earth System Model: A Framework for Collaborative Research},
  volume = {94},
  url = {https://journals.ametsoc.org/view/journals/bams/94/9/bams-d-12-00121.1.xml},
  year = {2013},
}

@misc{isabel2004,
  author       = {{National Center for Atmospheric Research}},
  title        = {Hurricane {Isabel} Simulation Dataset, {IEEE} Visualization 2004 Contest},
  year         = {2004},
  howpublished = {\url{http://vis.computer.org/vis2004contest/data.html}},
  note         = {WRF model simulation data produced by NCAR, courtesy of the U.S. National Science Foundation (NSF). Accessed: 2026-08-06}
}

@misc{pocket_agent_2024,
  author       = {{DIR-LAB}},
  title        = {Pocket-Agent},
  year         = {2024},
  publisher    = {GitHub},
  journal      = {GitHub repository},
  howpublished = {\url{https://github.com/DIR-LAB/pocket-agent}},
  note         = {Accessed: 2026-08-08}
}

@misc{anthropic2024mcp,
  author       = {{Anthropic}},
  title        = {Model Context Protocol},
  year         = {2024},
  howpublished = {\url{https://modelcontextprotocol.io/}},
  note         = {Accessed: 2026-08-08}
}

@misc{openai2025deepresearch,
  author       = {{OpenAI}},
  title        = {Introducing Deep Research},
  year         = {2025},
  month        = feb,
  day          = {2},
  howpublished = {\url{https://openai.com/index/introducing-deep-research/}},
  note         = {Accessed: 2025-03-20}
}

@article{bran2024chemcrow,
  title   = {Augmenting Large Language Models with Chemistry Tools},
  author  = {Bran, Andres M. and Cox, Sam and Schilter, Oliver
             and Baldassari, Carlo and White, Andrew D.
             and Schwaller, Philippe},
  journal = {Nature Machine Intelligence},
  volume  = {6},
  pages   = {525--535},
  year    = {2024},
  doi     = {10.1038/s42256-024-00832-8}
}

@inproceedings{Staller,
author = {Egersdoerfer, Chris and Carns, Philip and Snyder, Shane and Ross, Robert and Dai, Dong},
title = {STELLAR: Storage Tuning Engine Leveraging LLM Autonomous Reasoning for High Performance Parallel File Systems},
year = {2025},
isbn = {9798400714665},
publisher = {Association for Computing Machinery},
address = {New York, NY, USA},
url = {https://doi.org/10.1145/3712285.3759887},
doi = {10.1145/3712285.3759887},
booktitle = {Proceedings of the International Conference for High Performance Computing, Networking, Storage and Analysis},
pages = {1252–1266},
numpages = {15},
location = {
},
series = {SC '25}
}

@inproceedings{shinn2023reflexion,
  title     = {Reflexion: Language Agents with Verbal Reinforcement Learning},
  author    = {Shinn, Noah and Cassano, Federico and Berman, Edward
               and Gopinath, Ashwin and Narasimhan, Karthik R.
               and Yao, Shunyu},
  booktitle = {Advances in Neural Information Processing Systems},
  volume    = {36},
  year      = {2023}
}

@inproceedings{krasowska2021exploring,
  title     = {Exploring Lossy Compressibility through Statistical
               Correlations of Scientific Datasets},
  author    = {Krasowska, David and Bessac, Julie and Underwood, Robert
               and Calhoun, Jon C. and Di, Sheng and Cappello, Franck},
  booktitle = {2021 7th International Workshop on Data Analysis and
               Reduction for Big Scientific Data (DRBSD-7)},
  pages     = {47--53},
  year      = {2021},
  organization = {IEEE}
}

@misc{SZx,
  title         = {SZx: An Ultra-Fast Error-Bounded Lossy Compressor
                   for Scientific Datasets},
  author        = {Xiaodong Yu and Sheng Di and Kai Zhao and Jiannan Tian
                   and Dingwen Tao and Xin Liang and Franck Cappello},
  year          = {2022},
  eprint        = {2201.13020},
  archivePrefix = {arXiv},
  primaryClass  = {cs.DC}
}

@ARTICLE{8821342,
  author  = {Wang, Jinzhen and Liu, Tong and Liu, Qing and He, Xubin
             and Luo, Huizhang and He, Weiming},
  journal = {IEEE Transactions on Parallel and Distributed Systems},
  title   = {Compression Ratio Modeling and Estimation across Error Bounds
             for Lossy Compression},
  year    = {2020},
  volume  = {31},
  number  = {7},
  pages   = {1621--1635},
  doi     = {10.1109/TPDS.2019.2938503}
}

@inproceedings{jin2022improving,
  title     = {Improving Prediction-Based Lossy Compression Dramatically
               via Ratio-Quality Modeling},
  author    = {Jin, Sian and Di, Sheng and Tian, Jiannan and Byna, Suren
               and Tao, Dingwen and Cappello, Franck},
  booktitle = {2022 IEEE 38th International Conference on Data Engineering
               (ICDE)},
  pages     = {2494--2507},
  year      = {2022},
  organization = {IEEE}
}

@article{underwood2023black,
  title   = {Black-Box Statistical Prediction of Lossy Compression Ratios
             for Scientific Data},
  author  = {Underwood, Robert and Bessac, Julie and Krasowska, David
             and Calhoun, Jon C. and Di, Sheng and Cappello, Franck},
  journal = {The International Journal of High Performance Computing
             Applications},
  volume  = {37},
  number  = {3--4},
  pages   = {412--433},
  year    = {2023},
  publisher = {SAGE Publications}
}

@inproceedings{underwood2020fraz,
  author    = {Underwood, Robert and Di, Sheng and Calhoun, Jon C.
               and Cappello, Franck},
  title     = {{FRaZ}: A Generic High-Fidelity Fixed-Ratio Lossy
               Compression Framework for Scientific Floating-Point Data},
  booktitle = {2020 IEEE International Parallel and Distributed Processing
               Symposium (IPDPS)},
  pages     = {567--577},
  year      = {2020},
  publisher = {IEEE},
  doi       = {10.1109/IPDPS47924.2020.00065}
}

@article{underwood2022optzconfig,
  author  = {Underwood, Robert and Calhoun, Jon C. and Di, Sheng
             and Apon, Amy and Cappello, Franck},
  title   = {{OptZConfig}: Efficient Parallel Optimization of Lossy
             Compression Configuration},
  journal = {IEEE Transactions on Parallel and Distributed Systems},
  volume  = {33},
  number  = {12},
  pages   = {3505--3519},
  year    = {2022},
  doi     = {10.1109/TPDS.2022.3154096}
}

@inproceedings{mumenin2024qualitynet,
  title     = {QualityNet: Error-Bounded Lossy Compression Quality
               Prediction via Deep Surrogate},
  author    = {Mumenin, Khondoker Mirazul and Dai, Dong and Wang, Jinzhen
               and Di, Sheng},
  booktitle = {2024 IEEE International Conference on Big Data (BigData)},
  pages     = {252--261},
  year      = {2024},
  organization = {IEEE}
}

@article{Lindstrom2014,
  author    = {Peter Lindstrom},
  doi       = {10.1109/TVCG.2014.2346458},
  issn      = {10772626},
  number    = {12},
  journal   = {IEEE Transactions on Visualization and Computer Graphics},
  month     = dec,
  pages     = {2674--2683},
  publisher = {IEEE Computer Society},
  title     = {Fixed-Rate Compressed Floating-Point Arrays},
  volume    = {20},
  year      = {2014}
}

@INPROCEEDINGS{sperr,
  author    = {Li, Shaomeng and Lindstrom, Peter and Clyne, John},
  booktitle = {2023 IEEE International Parallel and Distributed Processing Symposium (IPDPS)},
  title     = {Lossy Scientific Data Compression With SPERR},
  year      = {2023},
  pages     = {1007--1017},
  doi       = {10.1109/IPDPS54959.2023.00104}
}
\end{document}